\documentclass[aps,prd,showpacs,
twocolumn
,superscriptaddress]{revtex4}
\usepackage{graphicx}
\usepackage{dcolumn}
\usepackage{bm}
\usepackage{color}
\usepackage[normalem]{ulem} 
\usepackage[dvipsnames]{xcolor} 
\usepackage{hyperref}
\hypersetup{
  colorlinks=true,        
  linkcolor=blue,         
  citecolor=cyan,         
}

\usepackage{doi}
\usepackage{amsmath}
\usepackage{amssymb}

\begin{document}

\title{Magnetized particle motion around 4-D Einstein-Gauss-Bonnet Black Hole}

\author{Javlon Rayimbaev}
\email{javlon@astrin.uz}
\affiliation{Ulugh Beg Astronomical Institute, Astronomicheskaya 33, Tashkent 100052, Uzbekistan}
\affiliation{National University of Uzbekistan, Tashkent 100174, Uzbekistan}
\affiliation{Institute of Nuclear Physics, Ulugbek 1, Tashkent 100214, Uzbekistan}

\author{Ahmadjon~Abdujabbarov}
\email{ahmadjon@astrin.uz}
\affiliation{Ulugh Beg Astronomical Institute, Astronomicheskaya 33, Tashkent 100052, Uzbekistan}
\affiliation{National University of Uzbekistan, Tashkent 100174, Uzbekistan}
\affiliation{Institute of Nuclear Physics, Ulugbek 1, Tashkent 100214, Uzbekistan}
\affiliation{Shanghai Astronomical Observatory, 80 Nandan Road, Shanghai 200030, P. R. China}
\affiliation{Tashkent Institute of Irrigation and Agricultural Mechanization Engineers, Kori Niyoziy, 39, Tashkent 100000, Uzbekistan}

\author{Bobur Turimov}
\email{bturimov@astrin.uz}
\affiliation{Ulugh Beg Astronomical Institute, Astronomicheskaya 33, Tashkent 100052, Uzbekistan}
\affiliation{Research Centre for Theoretical Physics and Astrophysics, Institute of Physics,\\
Silesian University in Opava, Bezru\v covo n\' am.13, CZ-74601 Opava, Czech Republic}

\author{Farruh Atamurotov}
\email{atamurotov@yahoo.com}
\affiliation{Inha University in Tashkent, Tashkent, 100107, Uzbekistan}
\affiliation{Ulugh Beg Astronomical Institute, Astronomicheskaya 33, Tashkent 100052, Uzbekistan}

\date{\today}

\begin{abstract}

In this paper, we have investigated the dynamics of magnetized particles around 4-D Einstein-Gauss-Bonnet black hole immersed in an external asymptotically uniform magnetic field. We have shown that the magnetic interaction parameter responsible for circular orbits decreases for negative values of the Gauss-Bonnet parameter $\alpha$ and the range where magnetized particle's stable circular orbits are allowed increases for the positive values of the parameter $\alpha$. The study of the collisions of magnetized, charged and neutral particles has shown that the center-of-mass energy of the particles increases in the presence of positive Gauss-Bonnet parameter. Finally, we show how the magnetic interaction and  Gauss-Bonnet parameter may mimic the effect of rotation of the Kerr black hole giving the same ISCO radius for magnetized particles. Detailed analysis of the ISCO show that spin of Kerr black hole can not be mimicked by the effects of magnetic interaction and the Gauss-Bonnet parameters when $\alpha<-4.37$ and the spin parameter $a > 0.237$. 

\end{abstract}

\maketitle

\section{Introduction}

Recent observations of gravitational waves~\cite{LIGO16} and black hole shadow~\cite{EHT19a} played the most important role of testing of general relativity in the strong field regime. However, the accuracy of those observations and experiments leave an open window for modified theories of gravity~\cite{LIGO16b, EHT19b}. The modification of general relativity can be considered as an approach to the unified theory of interactions. On the other hand, general relativity has a problem with explaining some observational data and singularity problems. These also force us to consider alternative and modified theories of gravity. 

Recently it was proposed a new approach of obtaining the four-dimensional solution of the Einstein-Gauss-Bonnet (EGB) gravity~\cite{Glavan20}. This approach has been obtained by avoiding Lovelock’s theorem which indicates that in 4-dimensional spacetime the general relativity with the cosmological constant is the only theory of gravity. However, authors of Ref~\cite{Glavan20} obtained 4-dimensional nonzero limits of EGB theory and its solution rescaling the Gauss-Bonnet term by the factor $1/(D-4)$, where D is the dimension of the spacetime.  

The proposed new or modified version of the theory has to be checked using observational and experimental data. Test of different gravity models using X-ray data from astrophysical objects has been considered in papers~\cite{Bambi12a, Bambi16b, Zhou18, Tripathi19}. The motion of the test or/and charged particles may also be used as a useful test of metric theory of gravity~\cite{Bambi17e, Chandrasekhar98}. Different properties of 4-D EGB black hole (BH) have been considered in~\cite{Malafarina20,Zhang20a,Aguilar19,Zhang20,
Guo20,Kumar20,Mishra20,Churilova20,Aragon20,Islam20}. The energetics and shadow of 6-D EGB BH have been studied in~\cite{Abdujabbarov15a}.  Here we plan to study the magnetized particle motion in the background spacetime of the 4-D EGB BH immersed in an external asymptotically uniform magnetic field.    

The magnetized particle motion around Shwarzshild and Kerr BHs immersed in an external magnetic field has been studied in~\cite{deFelice, defelice2004}. The magnetized particle motion around different types of compact objects has been studied in Refs~\cite{Rayimbaev16, Oteev16, Toshmatov15d, Abdujabbarov14, Rahimov11a, Rahimov11, Haydarov20} in the presence of the magnetic field. It is well known, that curved spacetime will change the uniform magnetic field structure~\cite{Wald74} and properties of the electromagnetic field and charged particles motion have been studied, e.g. in~\cite{Aliev02, Aliev2004, Aliev05, Aliev06, Frolov10, Abdujabbarov10, Abdujabbarov11a, Frolov12, Karas12a, Hakimov13, Stuchlik14a, Stuchlik16}. 

Another interesting approach to test the gravity theory is to consider the energetic processes around BH~\cite{Penrose69a, Blandford1977, Dhurandhar83, Dhurandhar84, Dhurandhar84b, wagh85, Banados09} which can be used to model the high energetic particle extraction from active galactic nuclei (AGNs).  In our previous works, we have considered the energetic processes around compact objects in different models of gravity~\cite{Abdujabbarov08,Abdujabbarov10, Abdujabbarov11a, Abdujabbarov13a, Abdujabbarov13b, Abdujabbarov14, Stuchlik14a, Narzilloev19, Oteev16, Benavides-Gallego18}.

In Ref~\cite{Banados09} authors have considered the particles acceleration around rotating Kerr BH. It was also shown that external magnetic filed around compact object may play a role of charged particles accelerator~\cite{Frolov12,Abdujabbarov13a}. Particularly, the acceleration mechanism for different scenarios have been considered in~\cite{Tursunov13,Tursunov16,Kolos15,Kolos17,Tursunov18a,Tursunov18,Shaymatov18,Kimura11,Zaslavskii12a,Banados11,Oteev16,Hakimov14a,Igata12,Toshmatov15d,Gao11,Liu11,Patil11b,Patil12,Patil10,
Zaslavskii11, Zaslavskii10b, Zaslavskii11c,Ghosh14, Abdujabbarov14,Shaymatov13}.

In this work, we plan to study magnetized particle motion and acceleration process around 4-D EGB  BH immersed in an external asymptotically uniform magnetic field. The paper organized as follow: In Sect~\ref{chapter1} we consider the electromagnetic field around 4-D EGB BH immersed in an external asymptotically uniform magnetic field and magnetized particles motion around it. The Sect~\ref{magaccelerat} is devoted to studying of the acceleration of the magnetized particles around 4-D EGB BH in the presence of the magnetic field. We analyze how the EBG gravity can mimic the spin of Kerr BH and magnetic interaction near the Schwarzschild BH in Sect~\ref{astroapll}. We conclude our results in Sect~\ref{Summary}. 

Throughout the paper, we use spacelike signature $(-,+,+,+)$ 
for the space-time and the unit system where $G = c = 1$ (However, for an astrophysical application we have written the speed of light explicitly in our expressions). The Latin indices run from $1$ to $3$ and the Greek ones from $0$ to $3$.

\section{Magnetized particles around 4-D EGB BH in external magnetic field\label{chapter1}}

In this section we will consider the dynamics of magnetized particles around a magnetized non-rotating BH in the background  of the D-dimensional EGB gravity, which describes by the action ~\cite{Glavan20}
\begin{eqnarray}
 {\cal S}=\int d^Dx\sqrt{-g}\left(2R+\alpha {\cal G}\right)\ ,
 \end{eqnarray} 
where $\alpha$ is Gauss-Bonnet (GB) coupling parameter, ${\cal G}$ is the Gauss Bonnet invariant which defines by the expression
\begin{eqnarray}
{\cal G}=R^{\mu \nu \sigma \rho}R_{\mu \nu \sigma \rho}-4R^{\mu \nu}R_{\mu \nu}+R^2\, .
\end{eqnarray}

The line element of the spacetime surrounding the non rotating BH in 4-D EGB theory has the following form~\cite{Glavan20}
\begin{eqnarray}\label{metric}
ds^2=-f(r)dt^2+f^{-1}(r)dr^2+r^2 d\theta^2 +r^2\sin^2\theta d\phi^2\ ,
\end{eqnarray}
where 
\begin{equation} \label{metfunct}
f(r)=1+\frac{r^2}{2\alpha}\left(1-\sqrt{1+\frac{8\alpha M}{r^3}} \right)\ . 
\end{equation}
At $\alpha \rightarrow 0$ the metric function $f(r)$ has the following limits 
\begin{eqnarray}
\lim_{\alpha \rightarrow 0} f(r)=1-\frac{2M}{r}\ . 
\end{eqnarray}

In this paper, we will use the metric function with $-$ sign which goes to Schwarzschild BH at zero the GB coupling parameter, $\alpha=0$. For simplicity of further calculations, in this work, we prefer to substitute $\alpha \to \alpha/M^2$. 

\subsection{4-D EGB BH immersed in external magnetic field}

We consider that the 4-D EGB BH is immersed in an external asymptotically uniform magnetic field and the expression for the electromagnetic four-potential around the BH can be found using the existence timelike and spacelike Killing vectors with the Wald method in the following form~\cite{Wald74}
\begin{eqnarray}\label{Aft}
A_{\phi} & = & \frac{1}{2}B_0 r^2\sin\theta,
\\\nonumber
A_t & = & 0 =A_r=A_{\theta}\ ,
\end{eqnarray}
where $B_0$ is the asymptotic value of the external magnetic field aligned along axis of symmetry, being perpendicular to the equatorial plane where $\theta =\pi/2$. 
The non zero components of the electromagnetic field tensor (${\cal F}_{\mu\nu}=A_{\nu,\mu}-A_{\mu,\nu}$) is
\begin{eqnarray}\label{FFFF}
{\cal F}_{r \phi}&=&B_0r\sin^2\theta
 \\
 {\cal F}_{\theta \phi}&=&B_0r^2\sin\theta \cos\theta\ .
\end{eqnarray}

One may now, calculate the orthonormal components of the magnetic field near the BH and non-zero components of the latter have the following form
\begin{equation}\label{BrBt}
    B^{\hat{r}}=B_0 \cos\theta, \\ \qquad B^{\hat{\theta}}=\sqrt{f(r)}B_0\sin \theta\ .
    \end{equation}
    One can see from Eq.(\ref{BrBt}) that only the angular component of the magetic field around the BH reflects the effects of strong gravity. However, the radial component of the magnetic field does not depend on it and has the same form as in Newtonian case.
    
\begin{figure}[h!]
    \centering
  \includegraphics[width=0.97\linewidth]{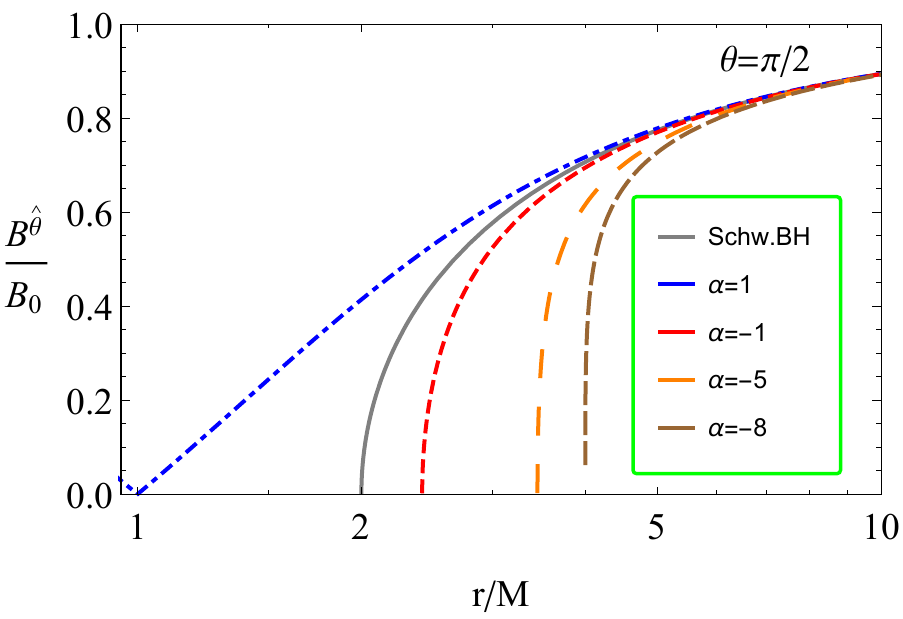}
    \caption{Radial profiles of normalized angular component of magnetic field $B^{\hat{\theta}}$ to  asymptotic value of the external magnetic field $B_0$.}
    \label{Bt}
\end{figure}

The radial dependence of the angular component of the magnetic field is shown in Fig.~\ref{Bt} for the different values of the GB parameter. From the dependence one can see that for positive values of $\alpha$ the angular component of the magnetic field increases with respect to Schwarzschild case. For the negative values of the GB parameter ($\alpha<0$) the magnetic field decreases and decays faster with decreasing radial coordinate. 
 
\subsection{The magnetized particle motion around 4-D EGB BH}
The motion of magnetized particles around a BH in the presence of the external magnetic field can be expressed using Hamilton Jacobi equation the following form \cite{deFelice}
\begin{eqnarray}\label{HJ}
g^{\mu \nu}\frac{\partial {\cal S}}{\partial x^{\mu}} \frac{\partial {\cal S}}{\partial x^{\nu}}=-\Bigg(m-\frac{1}{2} {\cal D}^{\mu \nu}{\cal F}_{\mu \nu}\Bigg)^2\ ,
\end{eqnarray}
where the term ${\cal D}^{\mu \nu}{\cal F}_{\mu \nu}$ is correspond to the interaction between the magnetized particle and the external magnetic field. In the case of, when consider the magnetic field of the magnetized particle has dipolar structure, the tensor ${\cal D}^{\alpha \beta}$ so called polarization tensor can be defined as
\begin{eqnarray}\label{dexp}
{\cal D}^{\alpha \beta}=\eta^{\alpha \beta \sigma \nu}u_{\sigma}\mu_{\nu}\ , \qquad {\cal D}^{\alpha \beta }u_{\beta}=0\ ,
\end{eqnarray} 
where $\mu^{\nu}$ is the four-vector of the magnetic dipole moment of the magnetized particle and $u^{\nu}$ - four velocity of the particle. The expression of the electromagnetic field tensor ${\cal F}_{\alpha \beta}$ with components of electric $E_{\alpha}$ and magnetic $B^{\alpha}$ fields has the following form
\begin{eqnarray}\label{fexp}
{\cal F}_{\alpha \beta}=u_{[\alpha}E_{\beta]}-\eta_{\alpha \beta \sigma \gamma}u^{\sigma}B^{\gamma}\ , 
\end{eqnarray}
where square brackets stand for antisymmetric tensor: $T_{[ab]}=\frac{1}{2}(T_{\mu \nu}-T_{\nu \mu})$ , $\eta_{\alpha \beta \sigma \gamma}$ is the pseudo-tensorial form of the Levi-Civita symbol $\epsilon_{\alpha \beta \sigma \gamma}$ with the relations 
\begin{eqnarray}
\eta_{\alpha \beta \sigma \gamma}=\sqrt{-g}\epsilon_{\alpha \beta \sigma \gamma}\, \qquad \eta^{\alpha \beta \sigma \gamma}=-\frac{1}{\sqrt{-g}}\epsilon^{\alpha \beta \sigma \gamma}\ ,
\end{eqnarray}
with $g={\rm det|g_{\mu \nu}|}=-r^4\sin^2\theta$ for metric (\ref{metric}) and 
\begin{eqnarray}
\epsilon_{\alpha \beta \sigma \gamma}=\begin{cases}
+1\ , \rm for\  even \ permutations
\\
-1\ , \rm for\  odd\  permutations
\\
0\ , \rm for\ the\ other\ combinations
\end{cases}\ .
\end{eqnarray}
One may easily calculate the interaction term ${\cal D}^{\mu \nu}{\cal F}_{\mu \nu}$, using the expressions (\ref{dexp}) and (\ref{fexp}) and we have, 
\begin{eqnarray}\label{DF1}
{\cal D}^{\mu \nu}{\cal F}_{\mu \nu}=2\mu^{\hat{\alpha}}B_{\hat{\alpha}}=2\mu B_0 {\cal L}[\lambda_{\hat{\alpha}}]\ ,
\end{eqnarray}
where $\mu$ is the module of the magnetic moment of the magnetized particle ($\mu = |\mu|=\sqrt{\mu_{\hat{i}}\mu^{\hat{i}}}$) and ${\cal L}[\lambda_{\hat{\alpha}}]$ is some characteristic function which defines the effect of comoving frame of reference with the particle rotating around the central object and it depends on the space coordinates, as well as other parameters defining the tetrad ${\lambda_{\hat{\alpha}}}$ attached to the comoving fiducial observer (i.e. the orbital angular velocity of the particle)~\cite{deFelice}.

Now we study the circular motion of the magnetized particle around the 4-D EGB BH in the weak magnetic interaction approximation (ether the external magnetic field weak enough or the particle less magnetized), with the limits of $\left({\cal D}^{\mu \nu}{\cal F}_{\mu \nu} \right)^2  \to 0$. We also consider that the magnetic moment of the particle is perpendicular to the equatorial plane where $\theta=\pi/2$, with the radial, angular, and azimuthal components  $\mu^{i}=(0,\mu^{\theta},0)$, respectively. The spacetime symmetries are preserved by the axial symmetric configuration of the magnetic field and, therefore, they still allow for two
conserved quantities: $p_{\phi}= L$ and $p_t = -E$ corresponding to angular momentum and energy
of the particle, respectively. So, we can write expression the action for the magnetized particle in the following form

\begin{eqnarray}\label{action}
{\cal S}=-E t+L\phi +{\cal S}_r(r)\ .
\end{eqnarray}
 The form of the action allows to separate variables in the Hamilton-Jacobi equation (\ref{HJ}).

One can easily get the expression for radial motion of the magnetized particle at the equatorial plane, with $p_{\theta}=0$, inserting Eq.(\ref{DF1}) to Eq.(\ref{HJ}) using the form of the action (\ref{action}) in the following form 
\begin{eqnarray}
\dot{r}^2={\cal{E}}^2-1-2V_{\rm eff}(r;\alpha ,l,\beta)\ ,
\end{eqnarray}
here ${\cal{E}}=E/m$ and $l=L/m$ are the specific energy and specific angular momentum of the particle, respectively, and the effective potential has the form
\begin{eqnarray}\label{effpot}
V_{\rm eff}(r;\alpha ,l,\beta)=\frac{1}{2}\Bigg[f(r)\left(1+\frac{l^2}{r^2}-\beta {\cal L}[\lambda_{\hat{\alpha}}]\right)-1\Bigg]\ ,
\end{eqnarray}
where the notation $\beta = 2\mu B_0/m$ is the magnetic coupling parameter which correspond to the interaction term ${\cal D}^{\mu \nu}{\cal F}_{\mu \nu}$ in the Hamilton-Jacobi equation (\ref{HJ}). 

The standard way to define the circular orbits of the particle around a central object is represented by the following conditions
\begin{eqnarray} \label{conditions}
\dot{r}=0 \ , \qquad \frac{\partial V_{\rm eff}}{\partial r}=0\ .
\end{eqnarray}
The first term of the condition (\ref{conditions}) and Eq.~(\ref{effpot})
give the following relation for the magnetic coupling parameter
\begin{eqnarray}\label{betafunc1}
\beta(r;l,{\cal E},\alpha)=\frac{1}{ {\cal L}[\lambda_{\hat{\alpha}}]}\Bigg(1+\frac{l^2}{r^2}-\frac{{\cal{E}}^2}{f(r)}\Bigg)\ .
\end{eqnarray}
The second part of the condition in Eq.(\ref{conditions}) satisfies the following relation
\begin{eqnarray}
 \frac{\partial V_{\rm eff}}{\partial r}=f(r) {\cal L}[\lambda_{\hat{\alpha}}] \frac{\partial \beta}{\partial r}\ ,
 \end{eqnarray}
is angular momentum of the magnetized particle rotating around the BH with spacetime (\ref{metric}).

Since, we consider the motion of the magnetic field at the equatorial plane, we interested in components of the magnetic field measured by the observer in a frame of comoving observer only in that plane and the components take the following form
\begin{eqnarray}\label{Bcomp}
B_{\hat{r}}=B_{\hat{\phi}}=0\ , \qquad B_{\hat{\theta}}=B_0f(r)\,e^{\Psi}\ ,
\\{\rm with}\, \qquad \label{epsi}
 e^{\Psi}=\left(f(r)-\Omega^2 r^2\right)^{-\frac{1}{2}}\ ,
\end{eqnarray}
where 
\begin{eqnarray}
  \Omega=\frac{d\phi}{d t}=\frac{d\phi/d\tau}{d t/d\tau}=\frac{f(r)}{r^2}\frac{l}{{\cal{E}}}\ .
\end{eqnarray}

Inserting Eq.(\ref{Bcomp}) into (\ref{DF1}) one can easily find the exact form of the interaction term of the Eq.~(\ref{HJ}) in the following form 
\begin{eqnarray}\label{DF2}
{\cal D}^{\mu \nu}{\cal F}_{\mu \nu}=2\mu B_0f(r)\,e^{\Psi}\ .\end{eqnarray}
The comparison of Eqs.(\ref{DF1}) and (\ref{DF2}) shown that the characteristic function has the form
\begin{eqnarray}\label{lambda}
{\cal L}[\lambda_{\hat{\alpha}}]=e^{\Psi}\, f(r)
\end{eqnarray}
Finally, one may find the exact form of the magnetic coupling $\beta(r;l,{\cal E},\alpha)$ parameter inserting Eq.s(\ref{lambda}) and (\ref{epsi}) in to Eq.(\ref{betafunc1}) in the following form 
\begin{eqnarray}\label{betafinal}
  \beta(r;l,{\cal E},\alpha)=\left(\frac{1}{f(r)}-\frac{l^2}{{\cal E}^2 r^2}\right)^{1/2}\Bigg(1+\frac{l^2}{r^2}-\frac{{\cal E}^2}{f(r)}\Bigg)\ .
\end{eqnarray}

The Eq.(\ref{betafinal}) has a physical meaning that a magnetized particle with specific energy ${\cal E}$ and angular momentum $l$ can be in the circular orbit in a given distance from the central BH, $r$.
 
\begin{figure}
  \centering
   \includegraphics[width=1\linewidth]{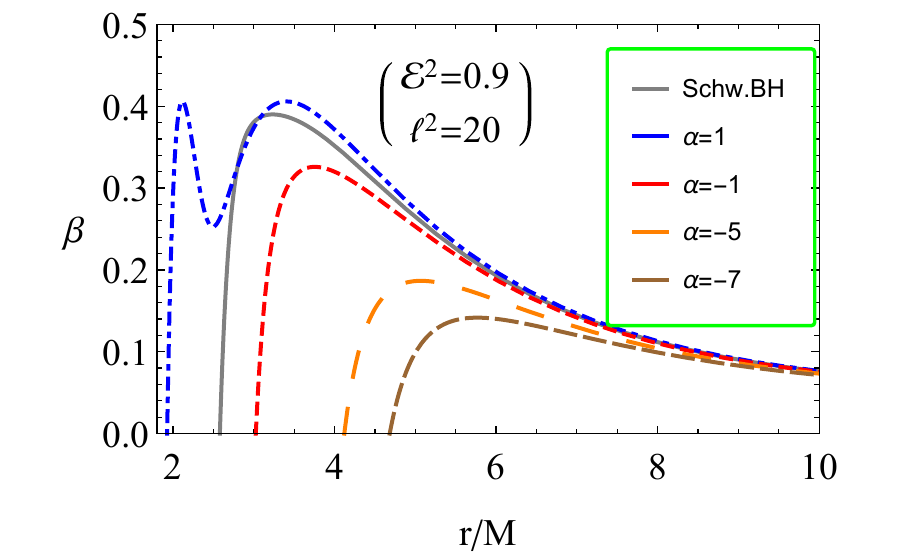}
   \includegraphics[width=1\linewidth]{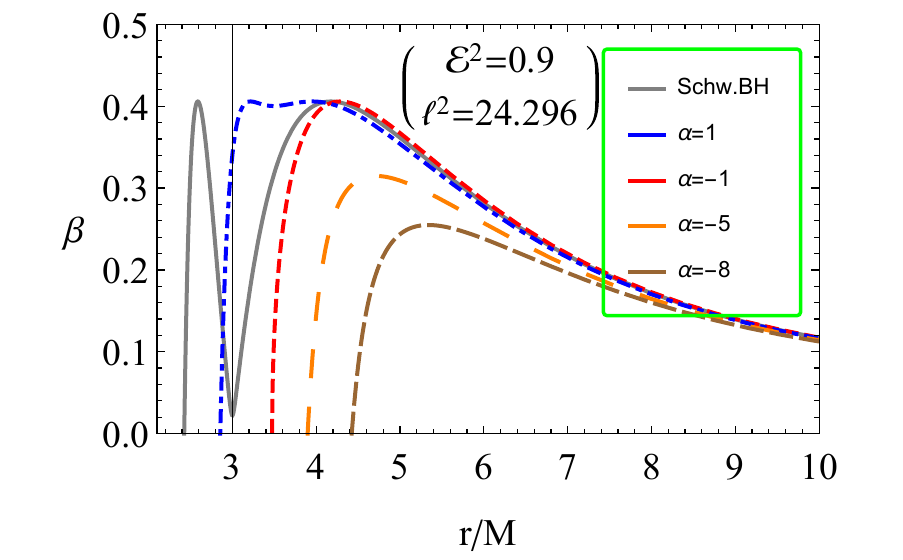}
   \caption{The radial dependence of magnetic coupling function for the different values of the Gauss-Bonnet coupling parameter, $\alpha$. The plots are shown for the different values of the specific angular momentum: $l^2=20$ (top panel) and $l^2=24.296$ (bottom panel) when ${\cal E}^2=0.9$. \label{betafig}}  
\end{figure}
  
The radial dependence of magnetic coupling function $\beta$ for the different values of $\alpha$ parameter is shown in Fig.~\ref{betafig}.  From the Fig.~\ref{betafig} one can see that the maximum value of the magnetic coupling parameter decreases (increases) and the distance where neutral particles $\beta=0$ can be in the circular orbits increases (decreases) with the increase of the absolute value of negative (positive) GB parameter, $\alpha$. Moreover, in case of when the Gauss-Bonnet parameter $\alpha=1$, the minimum of the magnetic coupling parameter, $\beta$ increases with the increase of specific angular momentum. However, the maximum value of the magnetic coupling parameter does not vary with increasing the value of the specific angular momentum, it implies that the effect of the GB coupling parameter stronger that the effect of centrifugal force acting on the magnetized particle.
  
Now, we will interest in the value of the magnetic coupling parameter for the stable circular orbits. The conditions for the stable circular orbits for magnetized particles in the following form
\begin{eqnarray}\label{conditionstab}
\beta =\beta(r;l,{\cal E},\alpha), \qquad \frac{\partial \beta(r;l,{\cal E},\alpha)}{\partial r}=0
\end{eqnarray}
Equation (\ref{conditionstab}) presents the system of equations with five variables $\beta,r,l,{\cal E}$ and the GB coupling parameter $\alpha$, so its solution can be parameterized in terms of any two of the five independent variables. For simplicity in our investigations we use the $\beta$ and the radius $r$ as free parameters. One may find the expressions for minimum value for the specific energy ${\cal E}$ of the magnetized  particle at stable circular orbit solving the equation $\partial \beta(r;l;{\cal E},\alpha)/\partial r=0$ with respect to the energy ${\cal E}$ and we have
\begin{eqnarray}\label{emin}
{\cal E}_{\rm min}(r;l,\alpha)=-\frac{l \left(1+\frac{8 \alpha  M}{r^3}-\left(1+\frac{2 \alpha }{r^2}\right) \sqrt{1+\frac{8 \alpha  M}{r^3}}\right)}{\sqrt{2 \alpha  \left(1+\frac{2 \alpha  M}{r^3}-\sqrt{1+\frac{8 \alpha  M}{r^3}}\right)}}
\end{eqnarray}

\begin{figure}
  \centering
   \includegraphics[width=0.95\linewidth]{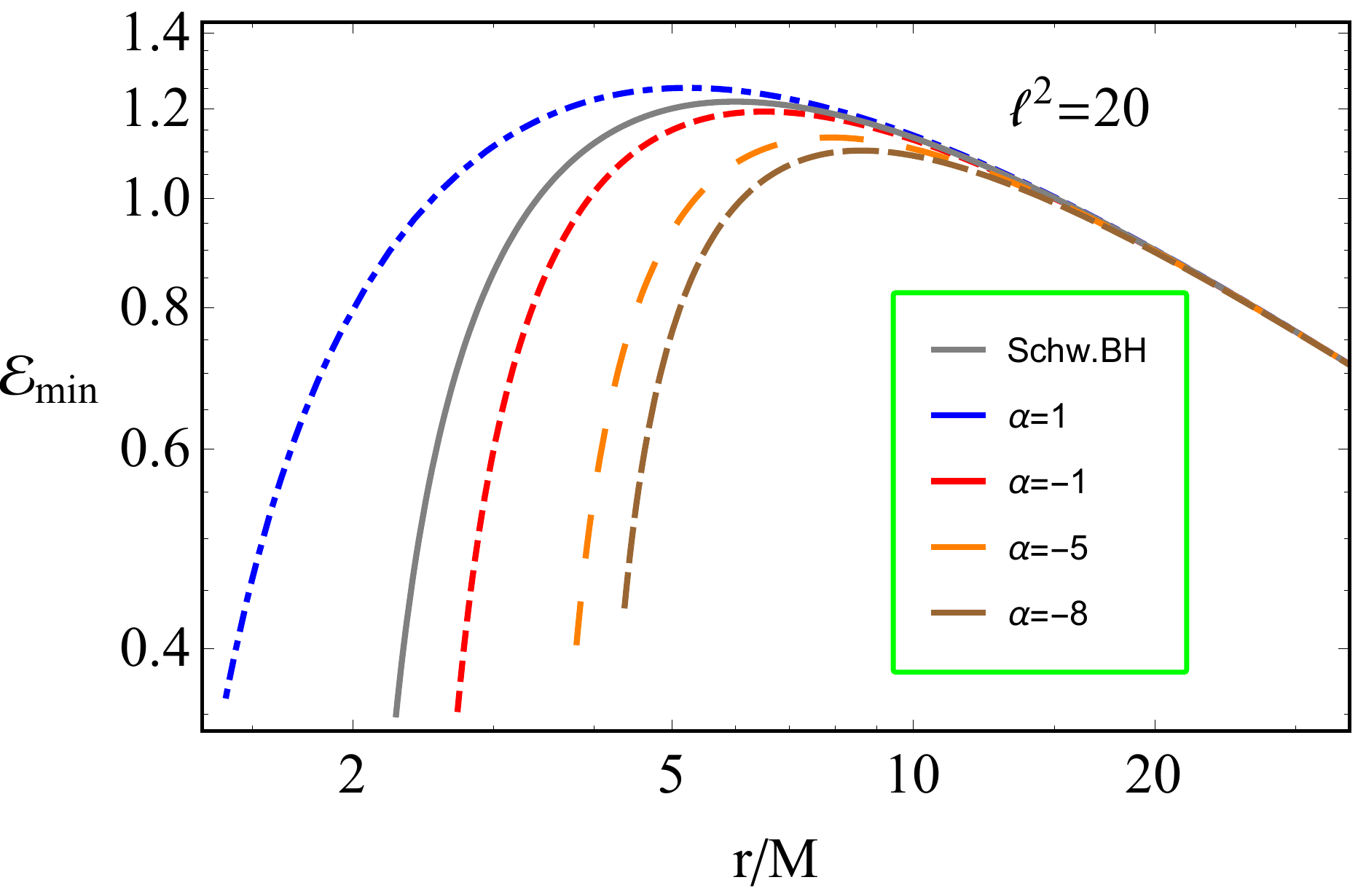}
      \caption{The radial dependence of the minimum value of specific energy of the magnetized particles for the different values of $\alpha$ parameter. \label{eminfig}}
\end{figure}

The radial dependence of the minimum value of specific energy of the magnetized particle is shown in Fig.~\ref{eminfig} for different values of the GB parameter $\alpha$. The maximum value of the specific energy increases as the increase of the value of the GB parameter and the distance where stable orbits are existed shifts to the outside from the central object due to the increase of the GB parameter.

One can easily obtain the expression for the minimum value of the magnetized coupling parameter $\beta$, substituting the expression (\ref{emin}) in to Eq.(\ref{betafinal}) and we have 

\begin{eqnarray}
\nonumber
\beta_{\rm min}(r;l,\alpha)=\frac{2 \alpha  \sqrt{1+\frac{8 \alpha  M}{r^3}-\frac{3 M}{r}\sqrt{1+\frac{8 \alpha  M}{r^3}}}}{r^2 \left(1+\frac{2 \alpha  M}{r^3}-\sqrt{1+\frac{8 \alpha  M}{r^3}}\right)}
\\
\frac{1+\frac{2 \alpha  M}{r^3}-\frac{6 \alpha  l^2 M}{r^5}-\left(1-\frac{2 \alpha  l^2}{r^4}\right) \sqrt{1+\frac{8 \alpha  M}{r^3}}}{\sqrt{1+\frac{8 \alpha  M}{r^3}} \left(1+\frac{2 \alpha }{r^2}-\sqrt{1+\frac{8 \alpha  M}{r^3}}\right)}
\end{eqnarray}

\begin{figure}
  \centering
   \includegraphics[width=0.98\linewidth]{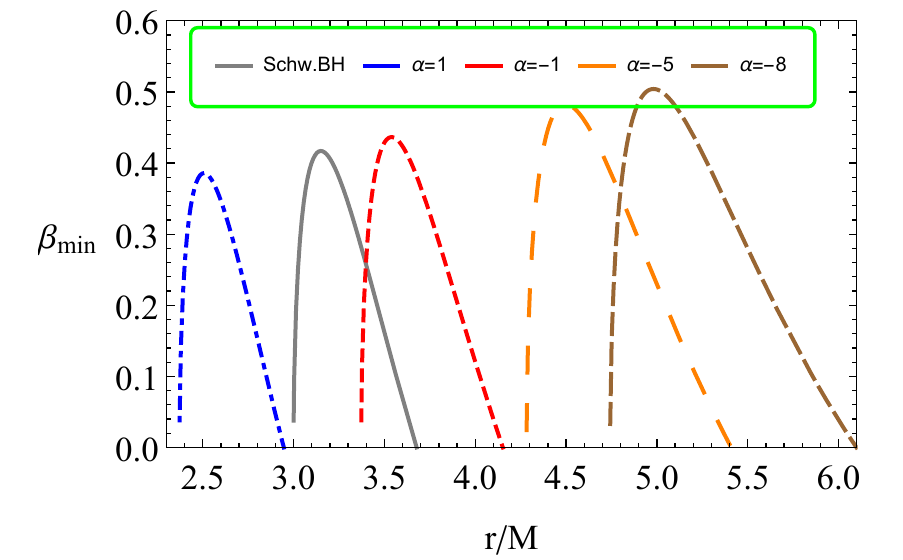}
    \caption{The radial dependence of the minimum value of the magnetic coupling parameter,$\beta$ for the different values of the GB parameter, $\alpha$. The plots are calculated at the value of the specific angular momentum $l^2=20$. \label{betaminfig}}
\end{figure}

Figure~\ref{betaminfig} shows the radial dependence of the minimal value of the magnetic coupling parameter of the magnetized particle for the different values of the $\alpha$ parameter. One can see from the figure that the maximal value of the minimum magnetic coupling parameter and the distance where it maximum increase with the increase of the GB parameter, $\alpha$.

Now we will look for the upper limit for stable circular orbits at some minimum value of the specific angular momentum. It corresponds to the extreme value of the minimum the magnetic coupling parameter and can be found solving the expression $\partial \beta_{\rm min}(r;l,\alpha)/\partial r=0$ concerning $l$ and we have

\begin{eqnarray}
\nonumber
&&l_{\rm min}(r; \alpha)=\frac{r^2\left( 1+\frac{2 \alpha  M}{r^3}-\sqrt{1+\frac{8 \alpha  M}{r^3}}\right)}{\sqrt{2 \alpha  \left(\sqrt{1+\frac{8 \alpha  M}{r^3}}-\frac{3 M}{r}\right)}}
\\
&&\times \left\{\left(3+\frac{4 \alpha}{r^2}\right)\sqrt{1+\frac{8 \alpha  M}{r^3}}-3\left(1+\frac{6\alpha  M}{r^3 }\right) \right\}^{-\frac{1}{2}}
\end{eqnarray}

\begin{figure}[ht!]
  \centering
   \includegraphics[width=0.98\linewidth]{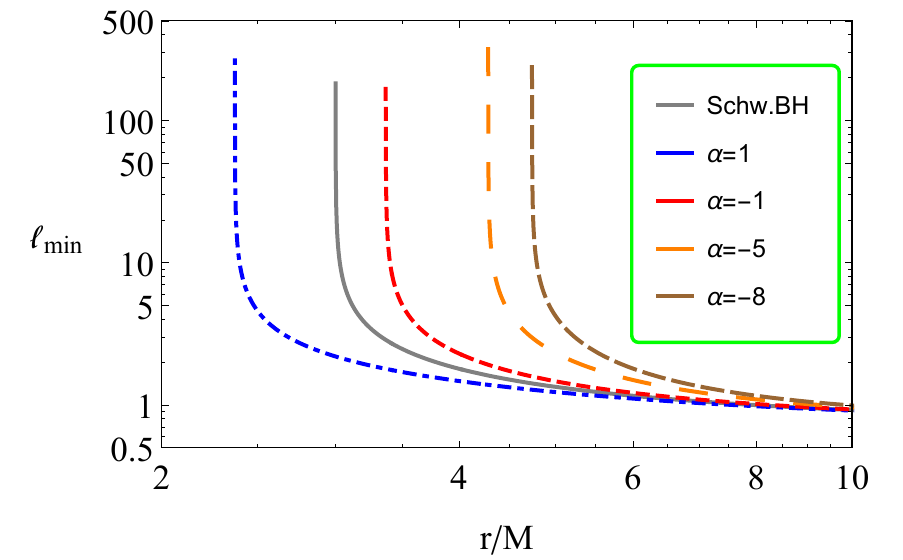}
    \caption{The radial dependence of minimal value of the specific angular momentum, $l_{\rm min}$, for the different values of the Gauss Bonnet parameter, $\alpha$. \label{lminfig}}
\end{figure}

Figure \ref{lminfig} shows the radial dependence of the minimum value of the specific angular momentum for the different values of the GB parameter, $\alpha$. One can see from the figure that the value of the minimum angular momentum as the increase of the GB parameter, $\alpha$. However, the distance where it is maximum decreases for the positive values of the parameter, $\alpha$.

\begin{eqnarray}
\beta_{\rm extr}(r;\alpha)=\frac{\sqrt{1+\frac{8 \alpha  M}{r^3}-\frac{3 M}{r}\sqrt{1+\frac{8 \alpha  M}{r^3}}}}{\left(1+\frac{3 r^2}{4 \alpha }\right) \sqrt{1+\frac{8 \alpha  M}{r^3}}-\frac{3}{2} \left(\frac{3 M}{r}+\frac{r^2}{2 \alpha }\right)}
\end{eqnarray}

\begin{figure}[h!]
  \centering
   \includegraphics[width=0.98\linewidth]{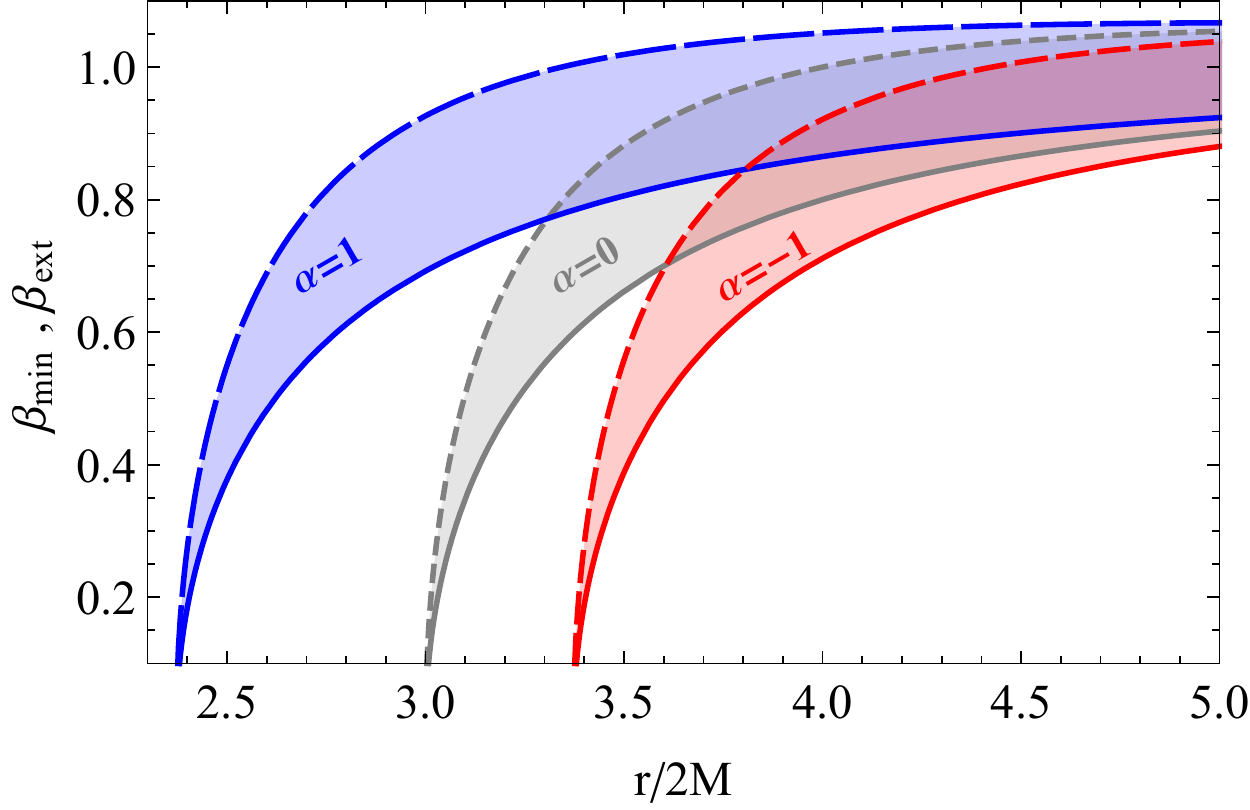}
    \caption{The radial dependence of the minimal value of magnetic coupling parameter at of the freely falling magnetized particle ($l=0$) and extreme values of the magnetized particles for the different values of the GB parameter, $\alpha$. Gray, light-blue and light-red colored areas correspond to the values of MOG parameter $\alpha=0$, $\alpha=0.1$ and $\alpha=0.2$, respectively.     \label{betaextremfig}}
\end{figure}

Figure \ref{betaextremfig} illustrates the radial dependence of the extreme value of the magnetic coupling parameter and the minimum value of the magnetic coupling parameter at $\beta_{\rm min}(l=0)$, for the different values of the GB parameter, $\alpha$. The colored areas imply the range where a magnetized particle with the magnetic coupling parameter $\beta_{\rm extr}<\beta<\beta_{\rm min}(l=0)$ stable circular orbits allowed. One can see from the figure that the minimum distance of the circular orbits increases with for the negative values of the GB  parameter $\alpha$. However, it is not possible to see the effect of the GB parameter $\alpha$, on the range where circular orbits are allowed. By the way, we show it in table making numerical calculations. 

\begin{table}[h!] \begin{center}\begin{tabular}{|c| c| c| c| c| c| c| }\hline
$\alpha$ & $\beta=0.1$ & $\beta=0.3$ & $\beta=0.5$ & $\beta=0.7$ & $\beta=0.9$ & $\beta=1$ \\[1.5ex]\hline \hline
$1 $ &4.46307 &  43.7175 & 146.911 & 413.763 & 9209.46 & $-$ \\[1.5ex]\hline
$0 $ & 4.20475 &  40,8344 & 134.883 & 371.121 & 11395 & $-$ \\[1.5ex]\hline
$-1 $ & 4.22257 &  40.8112 & 133.499 & 361.92 & 10149.97 & $-$\\[1.5ex]\hline
$ -5 $ & 4.28637 & 41.1339 & 132.961 & 352.387 & 11205.1 & $-$ \\[1.5ex] \hline
$-8$ & 4.2953 & 41.1471 & 132.31 & 349.839 & 1251.67& $-$ \\[1.5ex]\hline
\end{tabular} \end{center}
\caption{\label{tab} Numerical values for $\Delta r=r_{\rm max}-r_{\rm min}$ for the different values of the magnetic coupling parameter $\beta$ and the GB parameter $\alpha$ in the unit of $10^{-3}M$.}
 \end{table}

Table \ref{tab} demonstrates the range of circular orbits of the magnetized particle around the 4-D EGB BH for the different values of the magnetic coupling, $\beta$ and the GB parameter, $\alpha$. The values of the range $\Delta r$ given in the unit of $M/10^3$. One can see from the figure that the range $\Delta r$ increases with increasing the magnetic coupling parameter, $\beta$ and for positive values of the GB parameter, $\alpha$. However, the range decreases at $0> \alpha>-1$, then it increases again at $\alpha<-1$. 

\section{Magnetized particles acceleration near the 4-D EGB  BH \label{magaccelerat}}

It was first studied in Ref.~\cite{Banados09} that two particles falling to Kerr BH will be accelerated and in the case of extreme rotating BH the center of mass energy of colliding particles may diverge. The magnetic field may also play a role of charged particles accelerator~\cite{Frolov12, Abdujabbarov13a}. Here we will study the acceleration of magnetized particles near the 4-D EGB BH in the presence of the magnetic field. The center of mass energy of two colliding particles can be found by the relation
\begin{eqnarray}\label{ECMeq}
{\cal E}_{\rm cm}^2=\frac{E_{\rm cm}^2}{2m c^2}=1-g_{\alpha \beta}v_1^{\alpha}v_2^{\beta}\ ,
\end{eqnarray}
where $v_{i}^{\alpha}\ (i=1,2)$ are the 4-velocities of the colliding particles. Below we will consider several scenarios of collisions of (i) magnetized-magnetized, (ii) charged-magnetized, (iii) charged-charged and (iv) magnetized-neutral particles.

\subsection{Collision of two magnetized particles}

The four-velocity of the magnetized particle at equatorial plane ($\dot{\theta}=0$) has the following components:
\begin{eqnarray}\label{eqmotionm}
\nonumber
\dot{t}&=&\frac{{\cal E}}{f(r)}\ ,
\\
\nonumber
\dot{r}^2&=&{\cal E}^2-f(r)\left(1+\frac{l^2}{r^2}-\beta \right)\ ,
\\
\dot{\phi}&=&\frac{l}{r^2}\ .
\end{eqnarray}
Consequently the expression for center of mass energy of the two magnetized particles gets the following form:
\begin{eqnarray}
\nonumber
{\cal E}_{\rm cm}^2&=&1+\frac{{\cal E}_1{\cal E}_2}{f(r)}-\frac{l_1l_2}{r^2}-\\
\nonumber
&-&\sqrt{{\cal E}_1^2-f(r)\left(1+\frac{l_1^2}{r^2}-\beta_1\right)}
\\
&\times & \sqrt{{\cal E}_2^2-f(r)\left(1+\frac{l_2^2}{r^2}-\beta_2\right)}\ .
\end{eqnarray}

\begin{figure}
   \centering
  \includegraphics[width=0.98\linewidth]{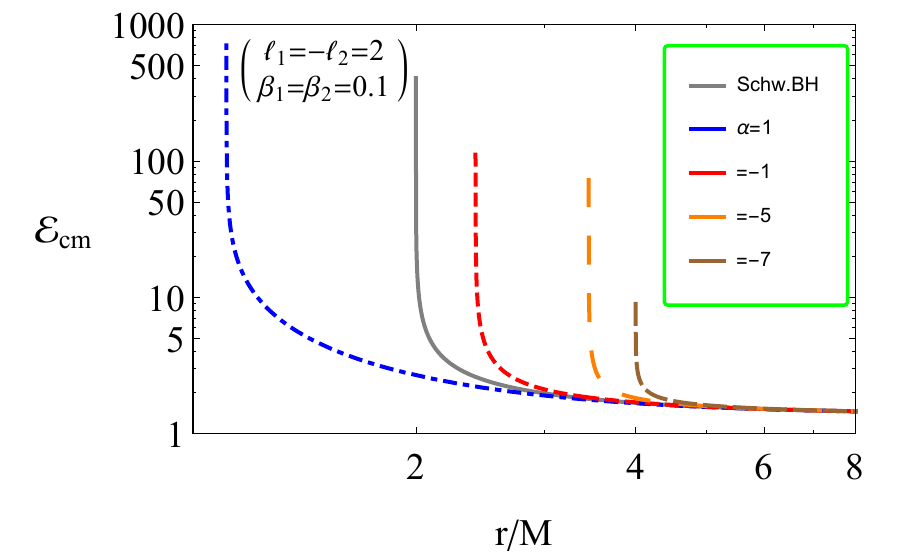}
	\caption{Radial dependence of the center of mass energy of collision of two magnetized particles with the same initial energy ${\cal E}_1={\cal E}_2=1$, around 4-D EGB BH for the different values of the $\alpha$ parameter.  \label{centermm}}
\end{figure}

Figure~\ref{centermm} illustrates the radial dependence of center-of-mass energy of magnetized particles for the different values of parameter $\alpha$. The plots are taken for the values of magnetized parameter $\beta_1=\beta_2=0.1$ and it was considered head-on collision of particles with the specific angular momentum $l_1=2, \, l_2=-2$. One can see from the Fig.~\ref{centermm} that the center-of-mass energy increases with the increase of the GB parameter $\alpha$. The distance where the center-of-mass energy reaches its maximum increases with the increase of the GB parameter $\alpha$. 

\subsection{Collision of two magnetized and charged particles}

In this subsection, we consider the collisions of the magnetized and charged particle. One can find the four-velocity of charged particle using the Lagrangian:
\begin{eqnarray}
{\cal L}=\frac{1}{2}mg_{\mu \nu}u^{\mu} u^{\nu}+e u^{\mu}A_{\mu} \ ,
\end{eqnarray}
here $e$ is electric charge of the particle. The conserved energy and angular momentum take the form:
\begin{eqnarray}
E&=& mg_{tt}\dot{t},
\\
L&=& mg_{\phi \phi}\dot{\phi}+eA_{\phi},
\end{eqnarray}
and one can find the expressions for the components of four velocity of the charged particle at equatorial plane in the following form:
\begin{eqnarray}\label{eqmotionch}
\nonumber
\dot{t}&=&\frac{{\cal E}}{f(r)}\ ,
\\
\nonumber
\dot{r}^2&=&{\cal E}^2-f(r)\Big[1+\Big(\frac{l}{r}-\omega_{\rm B} r\Big)^2\Big]\ ,
\\
\dot{\phi}&=&\frac{l}{r^2}-\omega_{\rm B}\ .
\end{eqnarray}
where $\omega_{\rm B}=eB/(2mc)$ is the cyclotron frequency which corresponding to interaction of magnetic field and charged particle.

Using Eqs.~(\ref{eqmotionch}), (\ref{eqmotionm}), and (\ref{ECMeq}) we may now write the expression for center-of-mass energy of magnetized and charged particles in the following form:
\begin{eqnarray}
\nonumber
{\cal E}_{\rm cm}^2&=&1+\frac{{\cal E}_1{\cal E}_2}{f(r)}-\left(\frac{l_1}{r^2}-\omega_{\rm B}\right)l_2\\
\nonumber
&-&\sqrt{{\cal E}_1^2-f(r)\Big[1+\Big(\frac{l_1}{r}-\omega_{\rm B} r\Big)^2\Big]}
\\
& \times & \sqrt{{\cal E}_2^2-f(r)\Big(1+\frac{l_2^2}{r^2}-\beta \Big)}\ .
\end{eqnarray}

\begin{figure}
    \centering
 \includegraphics[width=0.95\linewidth]{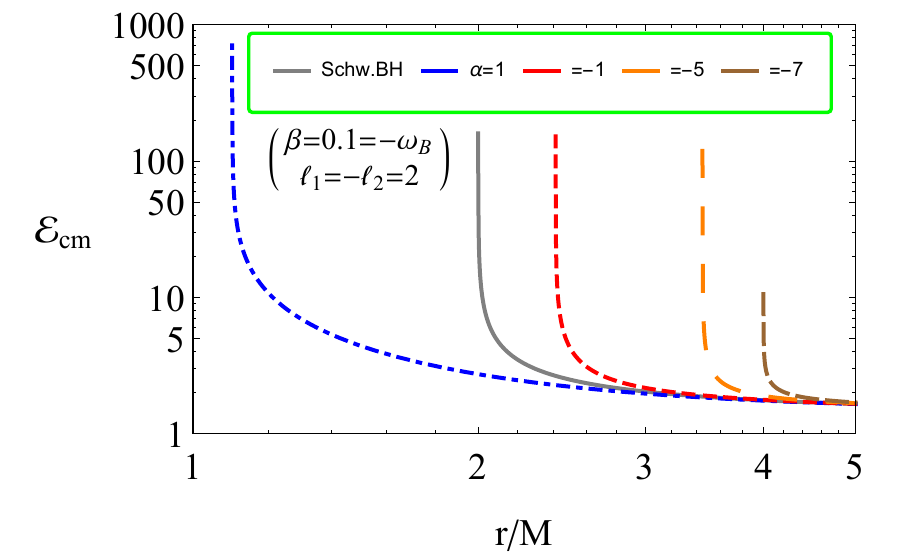}
 \includegraphics[width=0.95\linewidth]{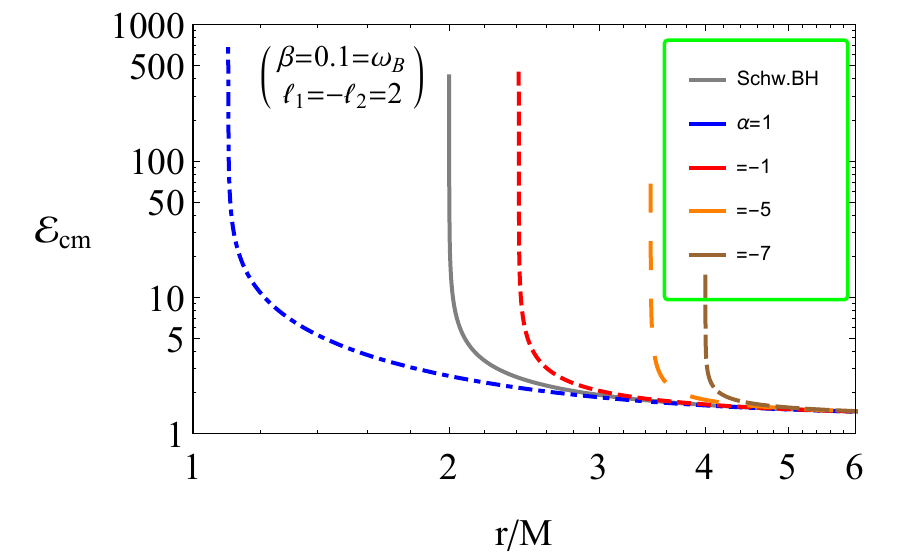}
   \caption{The radial dependence of the center-of-mass energy of collision of charged and magnetized particles with the energy ${\cal E}_1={\cal E}_2=1$, around 4-D EGB BH for the different values of the GB parameter$\alpha$. Top and bottom panels correspond to the cases of collisions of the magnetized particle with negatively and positively charged particles, respectively.      \label{centermmq}}
\end{figure}

Figure~\ref{centermmq} represents the radial dependence of center-of-mass energy of the colliding magnetized and charged particles around 4-D EGB BH. The values of specific angular momentum of the particles are chosen as $l_1=2,\,l_2=-2$ and dimensionless magnetic coupling and cyclotron parameters have the values $\beta=\omega=0.1$. One can see from both panels of the Fig.~\ref{centermmq} that at large distances the center-of-mass energy disappears due to repulsive Lorentz forces which means that at a large distance the collision doesn't happen. However, in the case of collision of magnetized and negative charged particle the distance where the center-of-mass energy disappears smaller than the case of the collision of a magnetized particle with positively charged particles due to the orientation of magnetic field: in the first case the Lorentz force has repulsive nature and in the second one attractive nature. Moreover, in both cases, the value of the center of mass energy increases with an increase of the $\alpha$ parameter.

\subsection{Collision of two magnetized and neutral particles}

We may now find the equations of motion for neutral particles around BH in EGB theory the following form:
\begin{eqnarray}\label{eqmotionn}
\nonumber
\dot{t}&=&\frac{{\cal E}}{f(r)}\ ,
\\
\nonumber
\dot{r}^2&=&{\cal E}^2-f(r)\Bigg(1+\frac{l^2}{r^2}\Bigg)\ ,
\\
\dot{\phi}&=&\frac{l}{r^2}\ .
\end{eqnarray}

Using the Eqs.~(\ref{eqmotionm}), (\ref{eqmotionn}), and (\ref{ECMeq}) the expression of center-of-mass energy of collision of neutral and magnetized particles can be written as:
\begin{eqnarray}\nonumber
{\cal E}_{\rm cm}^2=1+\frac{{\cal E}_1{\cal E}_2}{f(r)}-\frac{l_1l_2}{r^2}
&-&\sqrt{{\cal E}_1^2-f(r)\left(1+\frac{l_1^2}{r^2}-\beta \right)}
\\
& \times & \sqrt{{\cal E}_2^2-f(r)\left(1+\frac{l_2^2}{r^2}\right)}
\end{eqnarray}

\begin{figure}
    \centering
  \includegraphics[width=0.98\linewidth]{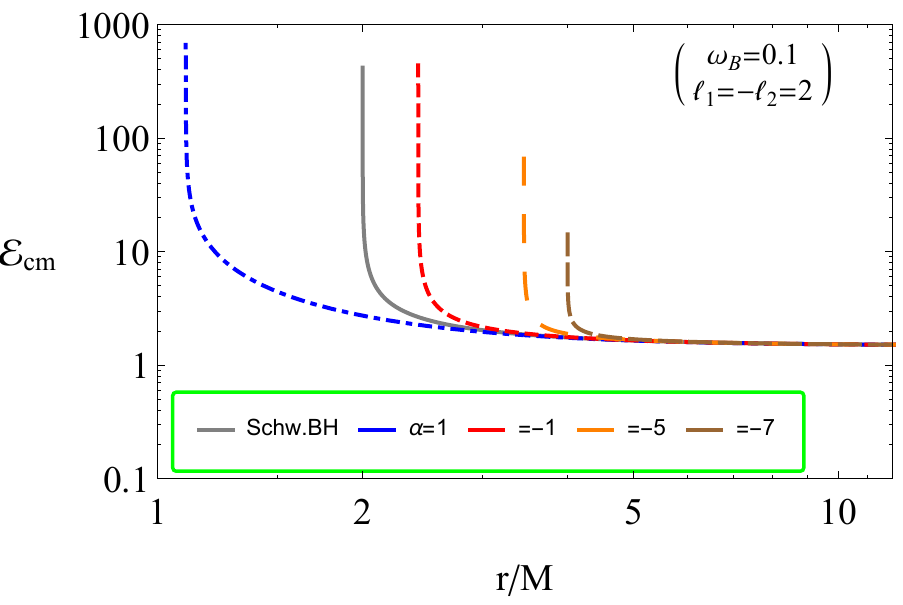}
      \caption{The radial dependence of the center-of-mass energy of collision of magnetized and neutral particles with the same initial energy ${\cal E}_1={\cal E}_2=1$, near EGB BH for the different values of $\alpha$. \label{centermn}}
\end{figure}

Figure~\ref{centermn} shows the radial dependence of center-of-mass energy of collision of neutral and magnetized particles with magnetic coupling parameter $\beta=0.1$ for the different values $\alpha$. In this case, one may also see that the increase of the parameter $\alpha$ causes the increase of the center-of-mass energy of the collision. However, the energy does not disappear due to the absence of Lorentz forces in the case of neutral particles. It implies that neutral particles can collide with other particles at any distance.

\subsection{Collision of two charged particles}

Finally, we consider the center of mass energy of collision of two charged particles and it can be found inserting Eq.(\ref{eqmotionch}) into (\ref{ECMeq})
\begin{eqnarray}
\nonumber
{\cal E}_{cm}^2=1&+ &\frac{{\cal E}_1{\cal E}_2}{f(r)}-r^2\left(\frac{l_1}{r^2}-\omega_{\rm B}^{(1)}\right)\left(\frac{l_2}{r^2}-\omega_{\rm B}^{(2)}\right)\\
\nonumber
&- &\sqrt{{\cal E}_1^2 -f(r)\Big[1+\Big(\frac{l_1}{r}-\omega_{\rm B}^{(1)} r\Big)^2\Big]}
\\
&\times &\sqrt{{\cal E}_2^2-f(r)\Big[1+\Big(\frac{l_2}{r}-\omega_{\rm B}^{(2)} r\Big)^2\Big]}\ .
\end{eqnarray}

\begin{figure*}
    \centering
   \includegraphics[width=0.48\linewidth]{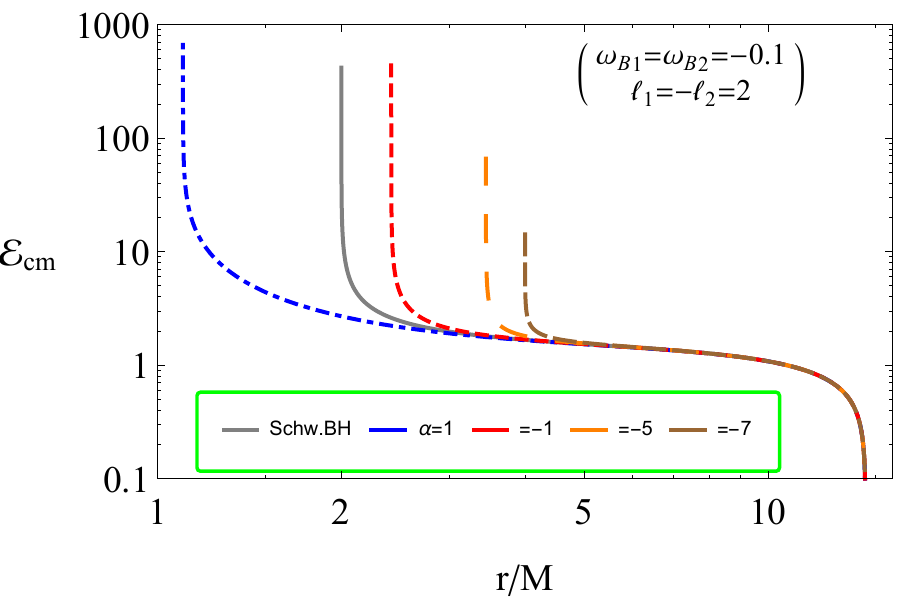}
   \includegraphics[width=0.48\linewidth]{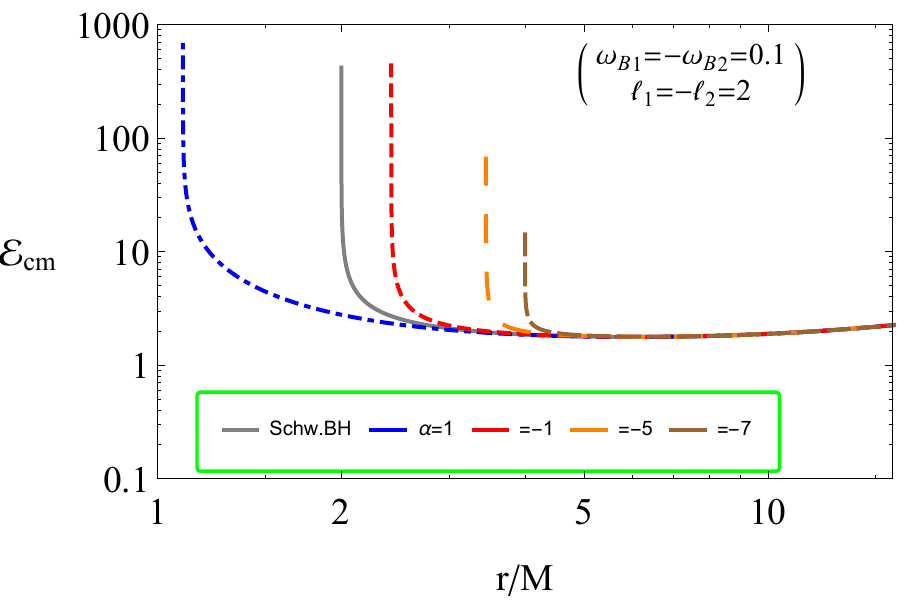}
  \includegraphics[width=0.48\linewidth]{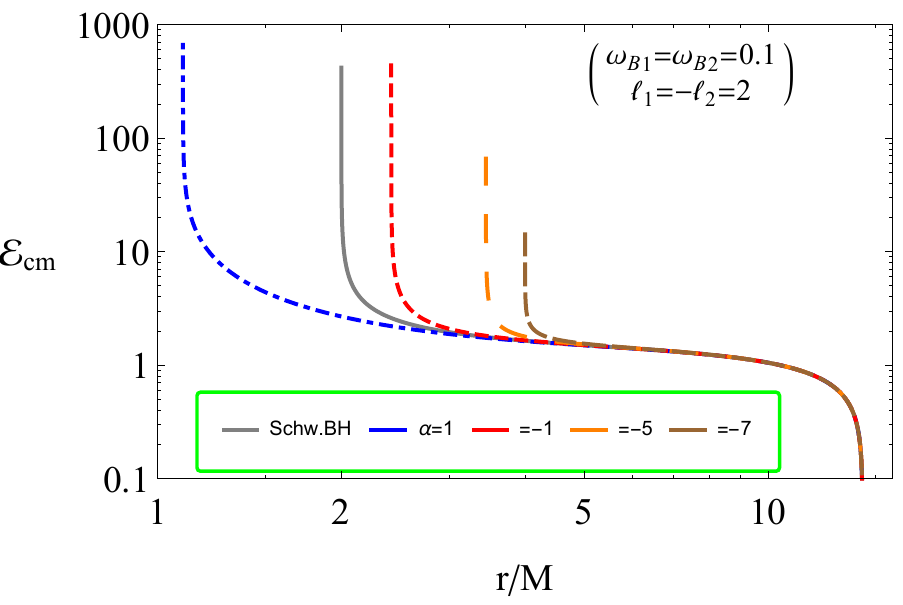}
   \caption{The radial dependence of the center of mass energy of collision of differently charged particles with the same initial energy ${\cal E}_1={\cal E}_2=1$, for the different values of $\alpha$, with the specific angular momentum $l_1=2$ and $l_2=-2$. Top-right, top left and bottom panels correspond to the cases of negative-negative, positive-positive and negative-positive charged particles collisions, respectively.\label{centermmch}}
\end{figure*}

Figure~\ref{centermmch} shows the radial dependence of the center-of-mass energy of charged particles collisions near BH in EGB theory for the different values of the $\alpha$ parameter. One can see from the figure in all cases the center-of-mass energy increases with the increase of the $\alpha$ parameter.  In the case of negative and positive charged particles the distance where the energy disappears increases with the increase of the value of the GB parameter $\alpha$ (bottom panel). Moreover, collisions of charged particles with the same sign show that the center-of-mass energy decreases again at the larger distances due to repulsive Coulomb interaction between the colliding particles and the collision does not take place at higher energies (top panels).

\section{Astrophysical applications \label{astroapll}}

Analysis of solution for ISCO radius of the magnetized particles around the 4-D  EGB BH in an external asymptotically uniform magnetic field and one for test particles around Kerr BH can help to find out how the GB coupling parameter mimics the effects of the rotation of BH giving the same value of  ISCO radius. In other words, our interest here is to show magnetic interaction mimicker of the 4-D EGB BH  parameter and/or spin of  BH.

ISCO radius of the test particles around Kerr BH for retrograde and prograde orbits can be expressed by the relations~\cite{Bardeen72}
\begin{eqnarray}
r_{\rm isco}= 3 + Z_2 \pm \sqrt{(3- Z_1)(3+ Z_1 +2 Z_2 )} \ ,
\end{eqnarray}
where
\begin{eqnarray} \nonumber
Z_1 &  = & 
1+\left( \sqrt[3]{1+ a}+ \sqrt[3]{1-a} \right) 
\sqrt[3]{1-a^2} \, ,
\\ \nonumber
Z_2 & = & \sqrt{3 a^2 + Z_1^2} \ .
\end{eqnarray}

In order to perform the above-mentioned task, we consider the motion of a magnetic particle in the following cases: (i) Kerr BH with rotation parameter $a$; (ii) 4-D  EGB BH (in the absence of magnetic field) with $\alpha$ parameter; (iii) 4-D EGB in the magnetic field and (iv) Schwarzschild BH in the magnetic field corresponding to magnetic interaction parameters in the range of $\beta \in (-0.5; 0.5)$.

\begin{figure*}
\centering
\includegraphics[width=0.8\linewidth]{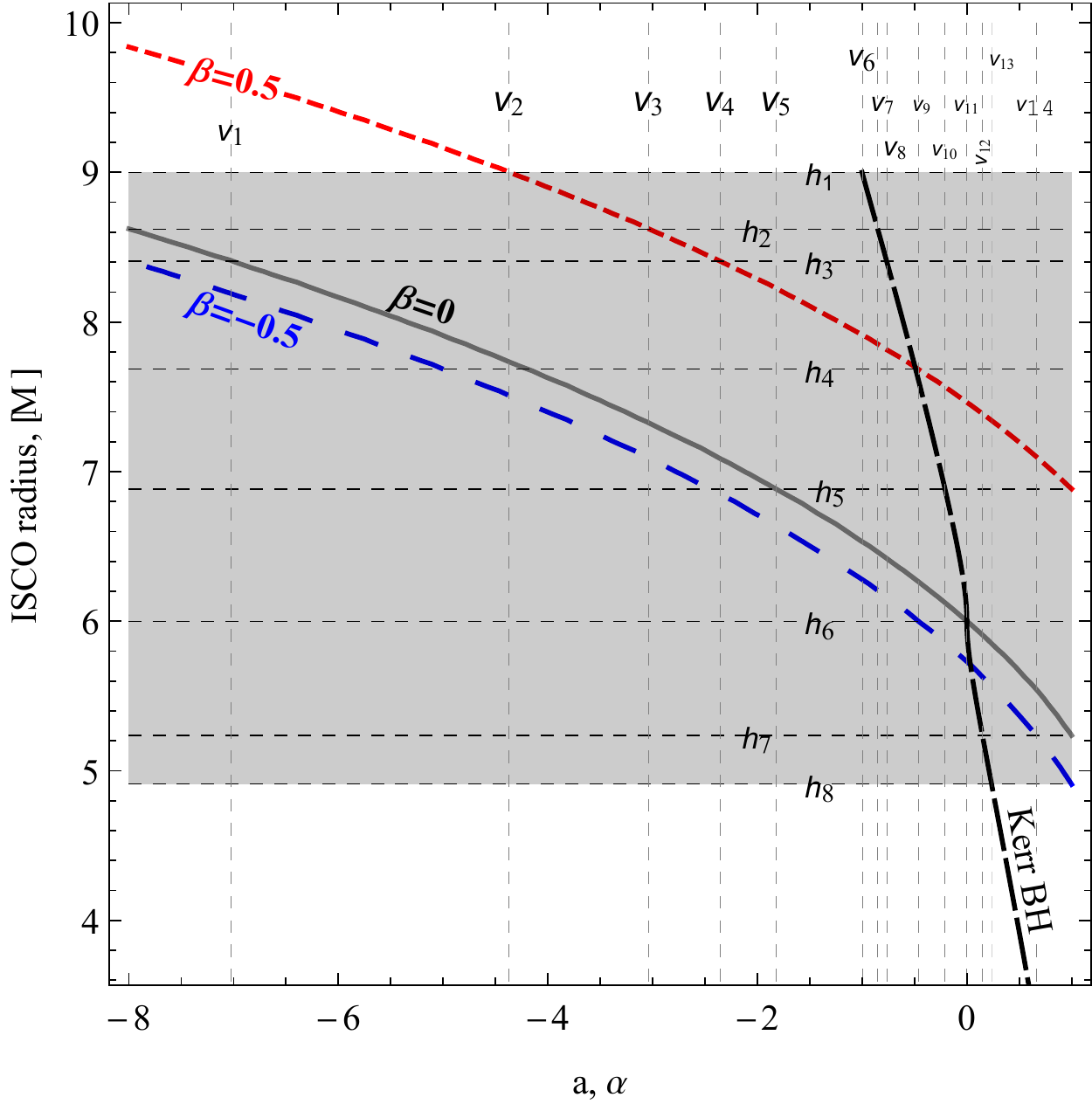}
\caption{The dependence of ISCO radius on spin and EGB coupling parameter. Black dashed and solid lines correspond to test particle ISCO around rotating Kerr BH and the 4-D EGB BH, respectively. Blue and red dashed lines correspond to ISCO of the magnetized particle around the 4-D EGB BH immersed in the magnetic field with the magnetic coupling parameters $\beta_1=-0.1$ and $\beta_2=0.1$, respectively. Vertical ($\rm v_i,\ i=1\div14$) and horizontal ($\rm h_i,\ i=1\div 8$) lines imply the important values for ISCO radius, rotation and EGB coupling parameters where the lines intersect. Spin and the coupling parameter $\alpha$ values are given in dimensionless unit as normalized to BH mass as $a/M$ and $\alpha/M^2$.} \label{iscoabeta}
\end{figure*}

ISCO radius of test particles around Kerr BH and magnetized particle around the 4-D EGB BH in magnetic fields as a function of spin and GB coupling parameter is shown in Fig.~\ref{iscoabeta}. Below we make analysis the plot of Fig.~\ref{iscoabeta} and how the one parameter can mimic other ones:  

\begin{enumerate}

\item {\bf Spin parameter of Kerr BH vs 4-D EGB parameter}

From the black large dashed and solid lines of the Fig.~\ref{iscoabeta} one can see that spin parameter of Kerr BH can mimic the 4D EBG gravity in the range of $\alpha \in (-8,1)$ with the values of  $a \in (-0.850707, 0.1464)$ (see the lines $\rm v_7$ and  $\rm v_{12}$ vertical lines) giving the same ISCO in the range of $r_{\rm isco} \in (5.236547,  8.62151)$ ($\rm h_2$ and $\rm h_7$ horizontal lines) and in the both cases when $a=0=\alpha$ the profiles of ISCO radius intersects at $r_{\rm isco}=6M$.

\item {\bf Spin parameter of Kerr BH vs 4-D EGB BH in magnetic field}

The spin parameter of Kerr BH with the values $a \in (-1, -0.208358 $ ($\rm v_7$ and $\rm  v_{10}$ vertical lines) can mimic the GB coupling parameter at $\alpha \in (-4.37059,1)$ (see the vertical line $\rm v_2$) giving the same  ISCO radius in the range of $r_{\rm isco} \in (6.8827, 9)$  (corresponding to the area between horizontal lines $\rm h_1$ and  $\rm h_5$) for magnetized particle with 
magnetic coupling parameter $\beta=0.5$. While the similar analysis for a magnetized particle with the parameter $\beta=-0.5$ show that the ISCO radius is same in the range of $r_{\rm isco} \in (5.2365, 8.4062)$ when the GB parameter $ 1 >\alpha > -8$ (areas corresponding between the horizontal lines  $\rm h_3$ and $\rm h_8$, and vertical lines $\rm v_8$ and $\rm v_{14}$). 

The cases when ISCO radius for the magnetized particles with the magnetic coupling parameter $\beta=0.5$ and $\beta=-0.5$ around the 4-D EGB BH in the absence and existence of the magnetic field, and rotating Kerr BH are the same at the range $6.8827\geq r_{\rm isco}\geq 8.4063$ (between $\rm h_3$ and $\rm h_5$ horizontal lines), for the value of the spin parameter $-0.850707\leq a\leq-0.20835$ ($\rm v_8$ and $\rm v_{10}$ vertical lines ) at the values of the GB coupling parameter: for the magnetized particle $\beta=0.5$ in the range $-2.35786 \leq \alpha \leq 1$, the magnetized particle with $\beta=-0.5$ in the range of the GB parameter $-8\leq \alpha\leq -2.35786$ (the vertical line $\rm v_4$) and for neutral particle (or vacuum around the 4-D EGB BH case) it is $-7.022\leq \alpha \leq -1.8202$ (the vertical lines $\rm v_1$ and $\rm v_5$).

\item {\bf Magnetic interaction vs 4-D EGB parameter}

In the case of magnetized particle motion around the 4-D EGB BH in the absence and presence of the external magnetic field, one can see the effect of the magnetic interaction at the values of the GB coupling parameter can mimic the magnetic coupling parameter $\beta=-0.5$ at the range of the GB parameter  $-7.022\leq \alpha\leq 0.6617$ (see $\rm v_1$ and $\rm v_{14}$ vertical and $\rm h_3$ and $\rm h_7$ horizontal lines) and $\beta=0.5$ at the range the parameter $1\leq \alpha \leq -0.8507$ giving the same value of ISCO radius (see $\rm v_3$ and $\rm v_5$ vertical and $\rm h_2$ and $\rm h_5$ horizontal lines).

\item {\bf Magnetic orientation vs 4-D EGB parameter}.

Consider the magnetized particle (with the values of magnetic interaction parameters $\beta=0.5$ and $\beta=-0.5$) motion around the 4-D EGB BH immersed in the uniform magnetic field. One can distinguish the cases of positive and negative values of the magnetic coupling parameter $\beta$ which is responsible to the orientation of the external magnetic field/dipole magnetic moment of the magnetized particle looking at the red and blue dashed lines between the horizontal lines $\rm h_3$ and $\rm h_5$. One can see from the range that it is possible to distinguish the positive and negative magnetic coupling parameters only if the GB parameter $\alpha \leq-2.35787$ for $\beta>0$ (the same direction with dipole magnetic moment of the magnetized particle) and $\alpha\geq -2.35787$ for $\beta<0$ (the opposite direction with dipole magnetic moment of the magnetized particle).

\end{enumerate}

\section{Summary and Discussions\label{Summary}}

In this work we have considered the motion of magnetized particles around the novel 4-D EGB BH immersed in an external asymptotically uniform magnetic field and have obtained the following main results

\begin{itemize}

\item The value of the magnetic coupling parameter responsible corresponding to the Lorentz force parallel to centrifugal one ($\sim l^2$) decreases (increases) in the presence of negative (positive) the GB parameter $\alpha$. This implies that the negative (positive) $ \alpha$ parameter decreases  (increases) the effective gravitational mass of the central object.

\item The analysis of stable circular orbits show that the range where stable circular orbits are allowed increases with the increase of the magnetic coupling parameter $\beta$. The parameter $\beta$ can not be stable at $\beta\geq1$. 

\item The studies of collisions of magnetized, neutral and charged particles show that the center-of-mass energy of the collisions increases with the increase of the GB parameter $\alpha$.

\item Finally, we have studied how the magnetic interaction and the 4-D EGB BH parameter can mimic the spin of Kerr BH. It was shown that one may distinguish the magnetized particle's ISCO with the magnetic coupling parameter around 4-D EGB BH and rotating Kerr BH only when the GB parameter $\alpha<-4.37059$ and the spin parameter $a>0.236593$.
\end{itemize}

\section*{Acknowledgement}

This research is supported by Grants No. VA-FA-F-2-008, No.MRB-AN-2019-29 of the Uzbekistan Ministry for Innovative Development. JR an AA thank Silesian University in Opava for the hospitality during their visit. 

\bibliographystyle{apsrev4-1}
\bibliography{gravreferences}

\end{document}